\documentclass[aps,prl,reprint,superscriptaddress]{revtex4-1}
\usepackage{amsmath}
\usepackage{graphicx}
\usepackage{units}  
\usepackage{xspace}
\usepackage{subfigure}
\usepackage{hyperref}

\newcommand{\ve}{\varepsilon}
\newcommand{\mb}{\mathbf}
\newcommand{\tb}{\textbf}
\newcommand{\beq}{\begin{equation}}
\newcommand{\eeq}{\end{equation}}
\newcommand{\bea}{\begin{eqnarray}}
\newcommand{\eea}{\end{eqnarray}}

\usepackage[usenames]{color}

\begin{document}

\bibliographystyle{apsrev}
 
\title{Chemical Potential Asymmetry and Quantum Oscillations in Insulators}

\date{\today}

\author{Hridis K. Pal}
\email{hridis.pal@physics.gatech.edu}
\affiliation{LPS, CNRS UMR 8502, Univ. Paris-Sud, Univ. Paris-Saclay, 91405  Orsay Cedex, France}
\author{Fr{\'e}d{\'e}ric Pi{\'e}chon}
\affiliation{LPS, CNRS UMR 8502, Univ. Paris-Sud, Univ. Paris-Saclay, 91405  Orsay Cedex, France}
\author{Jean-No{\"e}l Fuchs}
\affiliation{LPS, CNRS UMR 8502, Univ. Paris-Sud, Univ. Paris-Saclay, 91405  Orsay Cedex, France}
\affiliation{LPTMC, CNRS UMR 7600, Univ. Pierre et Marie Curie, 4 place Jussieu, 75252 Paris Cedex, France}
\author{Mark Goerbig}
\affiliation{LPS, CNRS UMR 8502, Univ. Paris-Sud, Univ. Paris-Saclay, 91405  Orsay Cedex, France}
\author{Gilles Montambaux}
\affiliation{LPS, CNRS UMR 8502, Univ. Paris-Sud, Univ. Paris-Saclay, 91405  Orsay Cedex, France}

\begin{abstract} 
We present a theory of quantum oscillations in insulators that are particle-hole symmetric and non-topological but with arbitrary band dispersion, at both zero and non-zero temperature. At temperatures $T$ less than or comparable to the gap, the dependence of oscillations on $T$ is markedly different from that in metals and depends crucially on the position of the chemical potential $\mu$ in the gap. If $\mu$ is in the middle of the gap, oscillations do not change with $T$; however, if $\mu$ is asymmetrically positioned in the gap, surprisingly, oscillations go to zero at a critical value of the inverse field determined by $T$ and $\mu$ and then change their phase by $\pi$ and grow again. Additionally, the temperature dependence is different for quantities derived from the grand canonical potential, such as magnetization and susceptibility, and those derived from the density of states, such as resistivity. However, the non-trivial features arising from asymmetric  $\mu$ are present in both.
\end{abstract}

\maketitle 

Quantum oscillations provide one of the most commonly used experimental tools to study metallic band structures, in both weakly and strongly correlated systems \cite{lif,sho,lut,eng,was1,was2}. They arise from Landau levels (LLs) crossing the Fermi level periodically as a function of the magnetic field. Such oscillations, therefore, are expected only in metallic systems with a Fermi surface.

Recently, this canonical understanding has been challenged by the observation \cite{tan} of quantum oscillations  in SmB$_6$ which is believed to be a topological Kondo insulator \cite{men,dze1,dze2,ale,lu}. While the exact origin of the oscillations is still being debated \cite{bas,ert}, it raises the questions: can quantum oscillations arise in insulators? If yes, how are they different from oscillations in metals? Two recent works have addressed these questions for specific models. Ref.~\cite{kno} considered a model--inspired by the experiment \cite{tan}--of a flat band hybridized with a dispersive band leading to a gap, and found oscillations in magnetization with a temperature dependence that is non-monotonic. However, it is not clear to what extent such findings depend on the enhanced density of states due to the flat band and the resulting strong particle-hole asymmetry. In contrast, Ref.~\cite{zha} considered a model of a topological insulator and found multiple phase changes in oscillations in density of states (DOS) accompanied by a non-monotonic temperature dependence. These features, however, are entirely a consequence of the topological properties of the model and are not expected in an ordinary insulator. A theory--and general understanding--of oscillations in insulators is missing.

In this Letter, we present a theory of quantum oscillations in insulators, at both zero and non-zero temperature. We construct our theory for a class of systems that are particle-hole symmetric and non-topological, but with arbitrary band dispersion. The motivation for adopting such a model is not  to simply contrast our results with those of Refs.~\cite{kno} and \cite{zha}: realistic systems with narrow gap and inverted bands where oscillations could be observed (reason for such requirements are discussed later), such as bilayer graphene at certain rotation angles \cite{mel,pal,bre}, materials at the onset of spin/charge density wave \cite{seb,car}, gapped nodal-line semimetals \cite{fu,bia}, etc., are, in fact, well described by the above model. In spite of the simplicity of our model, we find oscillations that do not follow the Lifshitz-Kosevich (LK) formula valid for metals, with features different from those reported in previous works \cite{kno,zha}.
The gap provides a new scale in the problem leading to new features when temperature $T$ is less than or comparable to this scale. Our main finding is that, in addition, the gap also provides a new degree of freedom not found in metals: the position of the chemical potential $\mu$ inside the gap. If $\mu$ is in the middle of the gap, i.e., $\mu=0$, oscillations do not change with temperature leading to a plateau in the temperature dependence; however, if it is asymmetrically positioned in the gap, i.e., $\mu\ne 0$ (but still in the gap), surprisingly, oscillations go to zero at a critical value of the inverse field determined by $T$ and $\mu$ and then change their phase by $\pi$ and grow again, mimicking properties of a topological insulator in an ordinary insulator! Additionally, oscillations behave differently for physical observables that are derived from the grand canonical potential and those that are related to the DOS; however, the non-trivial features arising from $\mu\ne 0$ are present in both families of observables.

\begin{figure}
\centering
\subfigure[]{\includegraphics[width=.28\columnwidth]{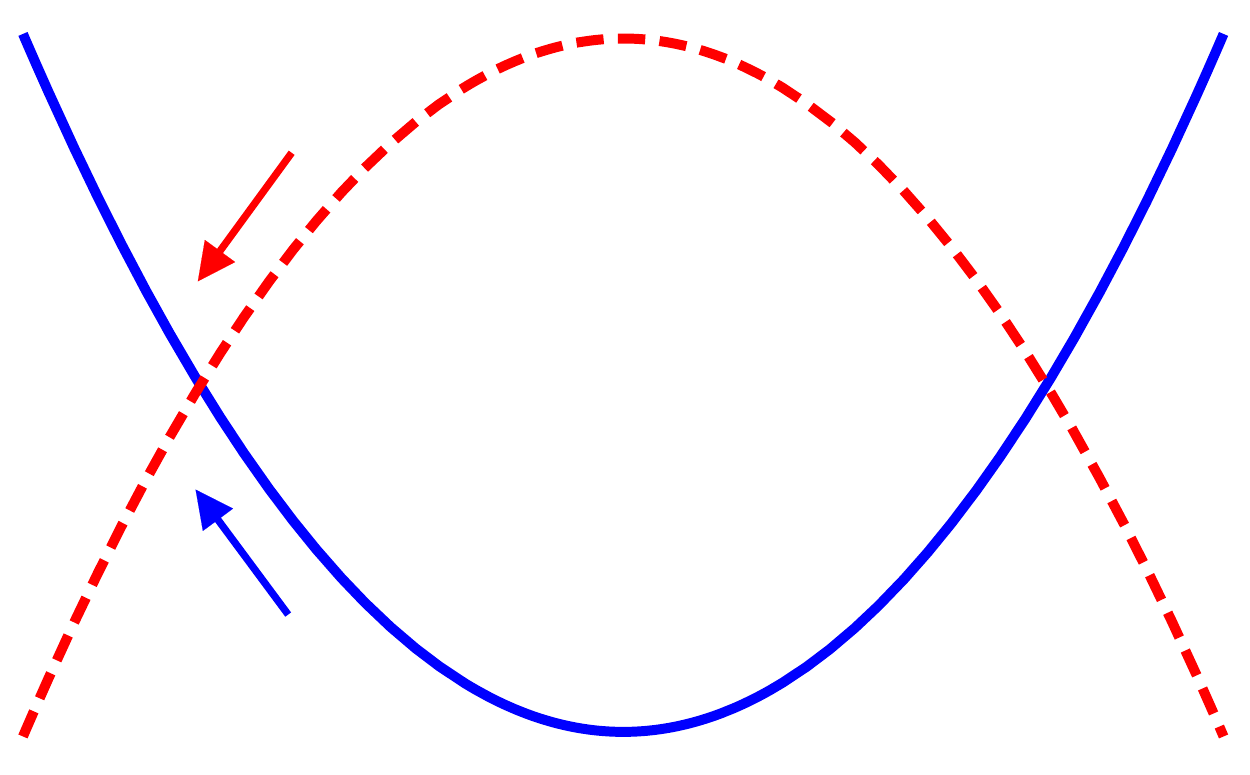}
\label{fig11}}
\quad
\subfigure[]{\includegraphics[width=.28\columnwidth]{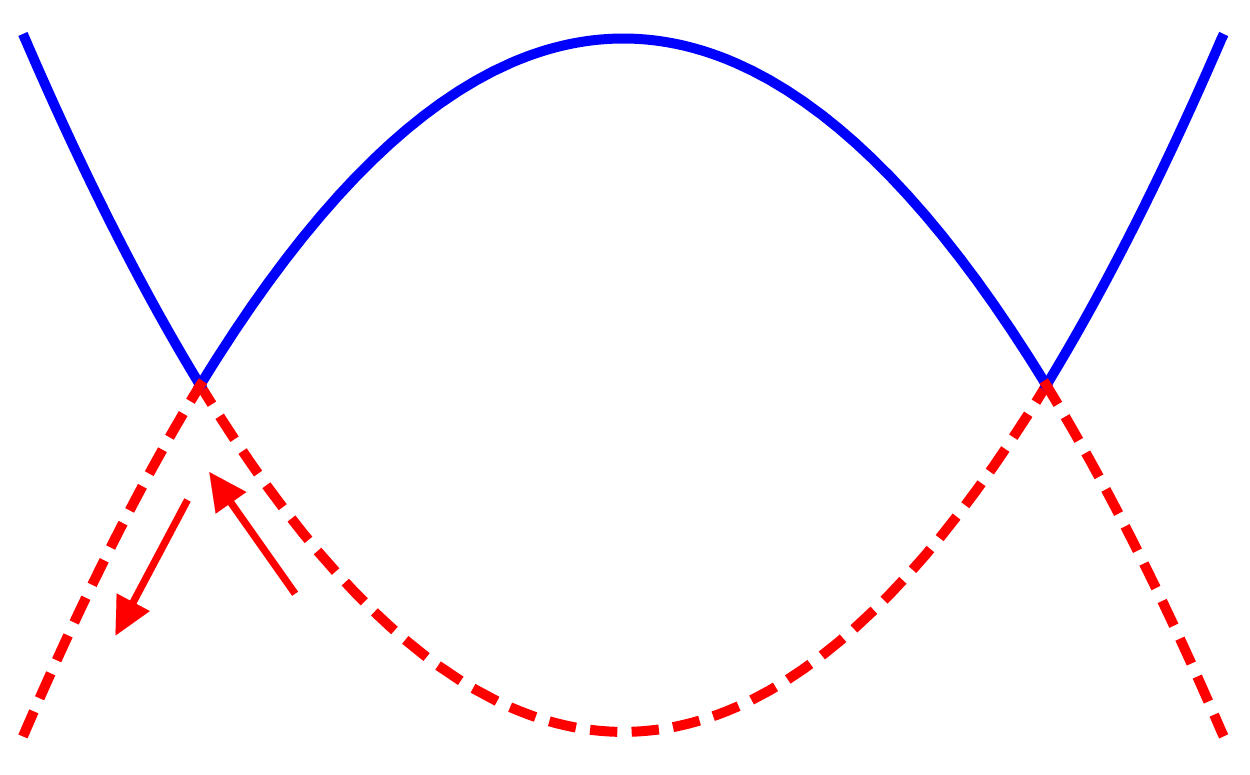}
\label{fig12}}
\quad
\subfigure[]{\includegraphics[width=.28\columnwidth]{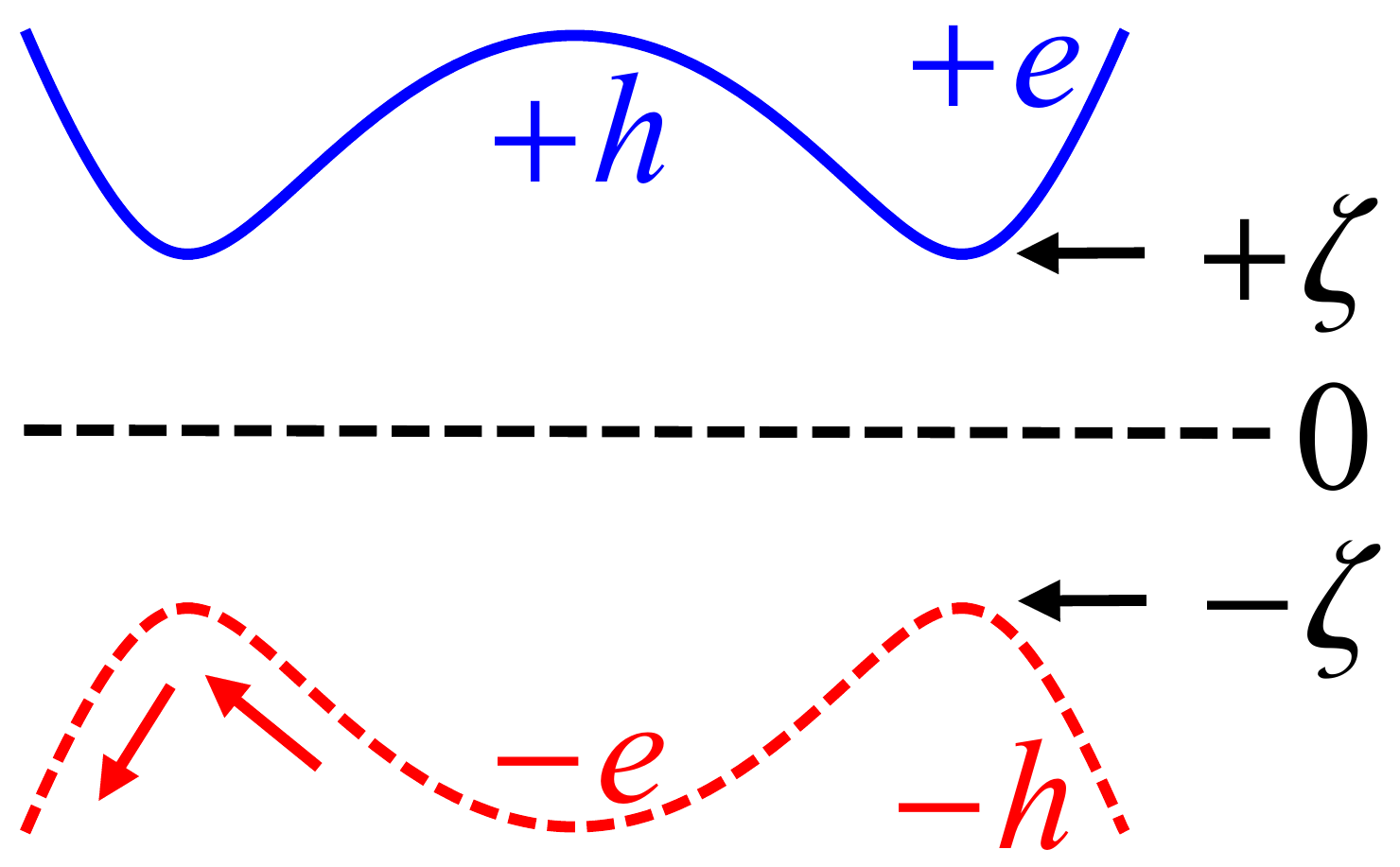}
\label{fig13}}
\caption{Alternative ways of viewing overlapping bands: in terms of (a) electron (blue, solid) and hole (red, dashed) levels, and (b) occupied (red, dashed) and unoccupied (blue, solid) levels. (c) The latter picture easily extends to the case when a gap opens  (gap exaggerated for clarity): $\pm$ are band indices and $e/h$ refer to electron/hole-like parts. In the presence of a magnetic field LLs are formed. The arrows show the direction in which they move as field is increased.}
\label{fig1}
\end{figure} 

Consider two identical overlapping bands with opposite curvature hybridized by some parameter. In band space, the Hamiltonian can be written as ($\hbar=k_B=1$)
\beq
H_{\mb{k}}=
\begin{pmatrix}
\varepsilon_{\mb{k}}-\Delta&\zeta\\
\zeta&-\varepsilon_{\mb{k}}+\Delta
\end{pmatrix},
\label{ham}
\eeq
where $\varepsilon_{\mb{k}}$ and $2\Delta$ denote the band dispersion (assumed non-topological)  and band overlap in the absence of the gap, respectively, and $\zeta>0$ is  a parameter that opens a gap (we assume $\varepsilon_{k=0}=0$ is an extremum and $\Delta$ has the appropriate sign to ensure band overlap at $\zeta=0$). The LLs for the Hamiltonian in Eq.~(\ref{ham}) are given by:
\beq
E_n^{\pm}=\pm\sqrt{(\varepsilon_n-\Delta)^2+\zeta^2},
\label{landaulevel}
\eeq
where $\varepsilon_n$ denotes the LLs corresponding to  $\varepsilon_{\mb{k}}$. 
In the ungapped case, the customary way to understand quantum oscillations is in terms of electron and hole levels in the two bands crossing the Fermi level $\mu$ at the intersection of the bands in opposite directions as a function of the magnetic field [Fig.~\ref{fig1}(a)]. An alternative way is to think only in terms of occupied levels in the two bands. The LLs from the electron band, on reaching the Fermi level, just `rolls over' to the hole band [arrows in Fig.~\ref{fig1}(b)]. 
As seen in Fig.~\ref{fig1}(c), even when the system is gapped, one can still separate the lower filled band into electron-like and hole-like parts. The band edge  $E_v=-\zeta$ now plays a role similar to the Fermi level in the ungapped case, giving rise to oscillations \cite{zha}.

\emph{Zero temperature.}---Consider a 2D system described by the Hamiltonian in (\ref{ham}). In the presence of a magnetic field $B$ the grand canonical potential at $T=0$ reads: $\Omega=D\sum_{E_n^{-}\le E_v}[E_n^{-}-E_v]$,
where $D=geB/2\pi$ is the degeneracy factor in 2D ($g$ denotes any extra degeneracy from internal degrees of freedom) \cite{sho}.
Following our previous discussion, we decompose the sum into two parts:
\beq
\frac{\Omega(B)}{D}=\sum_{n=0}^{N_v}[E_n^{-,e}-E_v]+\sum_{N_v+1}^{\Lambda}[E_n^{-,h}-E_v]=\frac{\Omega^{-,e}}{D}+\frac{\Omega^{-,h}}{D},
\label{omegaeh}
\eeq
where $E_n^{-,e/h}$ denote the electron-like and hole-like parts of the lower filled band, respectively, $N_v$ is the highest LL with energy smaller than $E_v$ in the electron-like part of the spectrum, and $\Lambda$ is a cutoff for the hole-like part of the band ($E^{-}_{\Lambda}\sim$ bandwidth) \cite{com1}.

\begin{figure}
\centering
\subfigure[]{\includegraphics[width=.45\columnwidth]{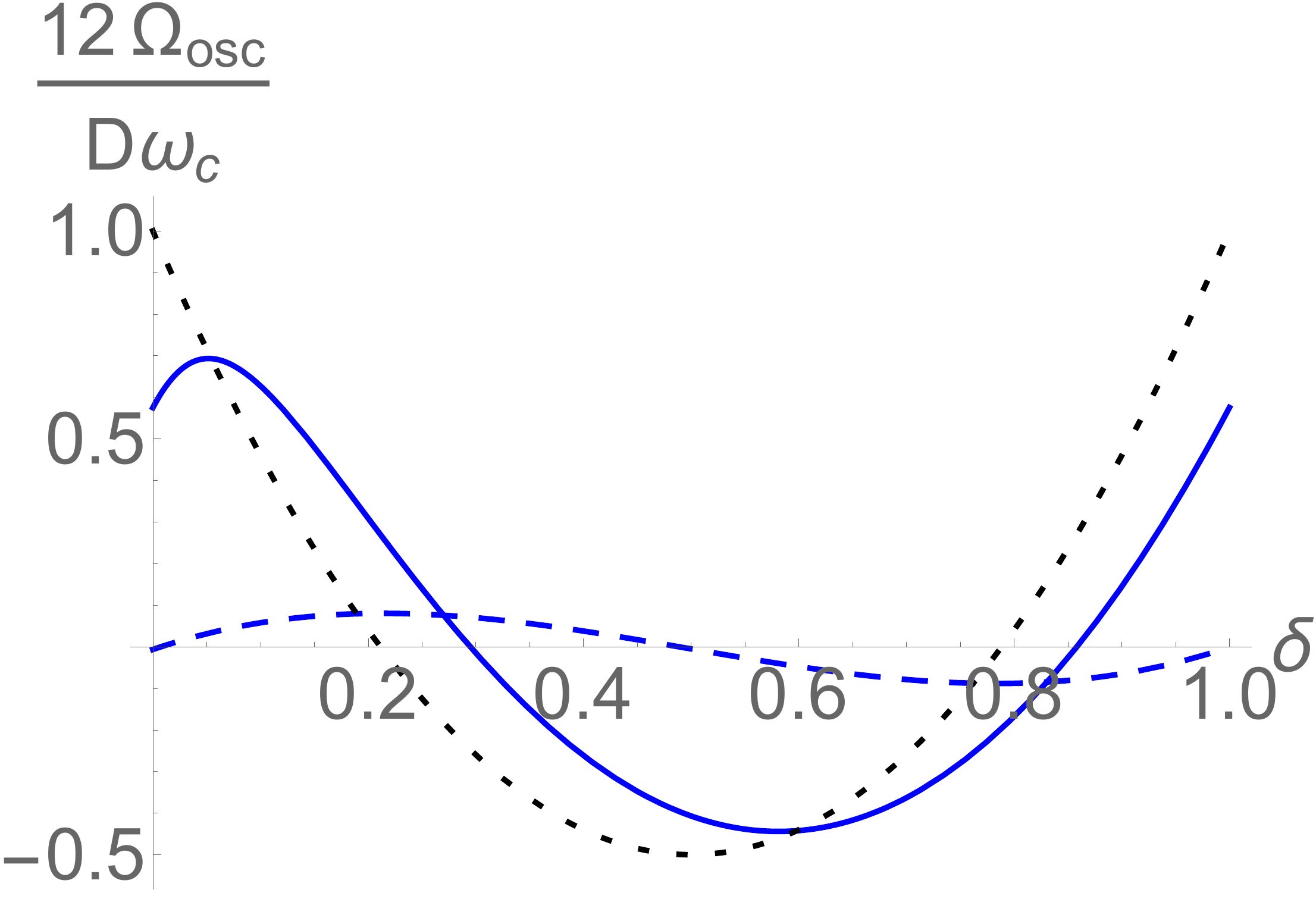}}
\quad
\subfigure[]{\includegraphics[width=.45\columnwidth]{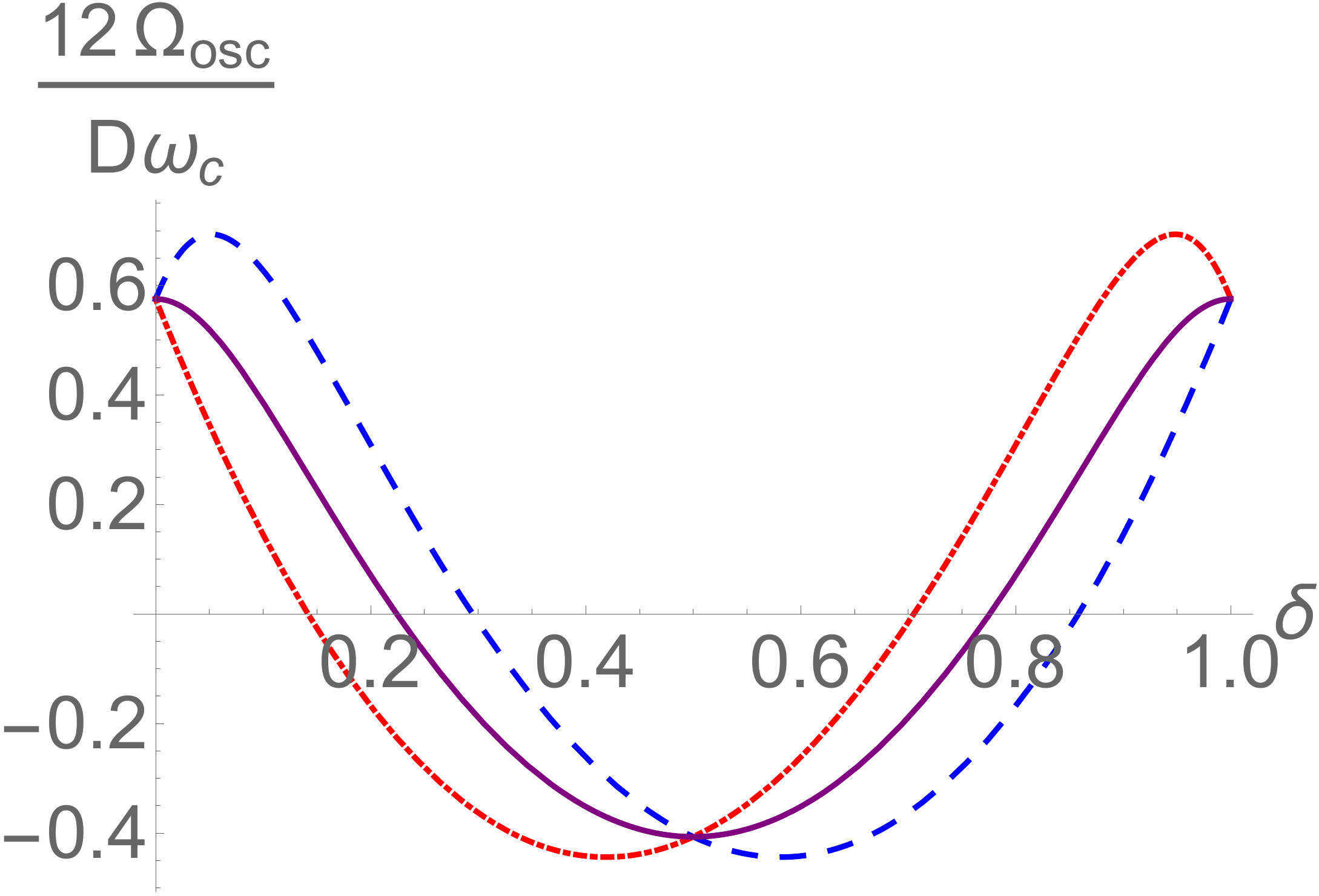}}
\caption{(a) $\Omega^{-,e}_{osc}$ for $\zeta/\omega_c=0.1$ (blue, solid) and $\zeta/\omega_c=1.2$ (blue, dashed) compared with the case of $\zeta=0$ (black, dotted). (b) $\Omega^{-,e}_{osc}$ (blue, dashed), $\Omega^{-,h}_{osc}$ (red, dotdashed), and $(\Omega^{-,e}_{osc}+\Omega^{-,h}_{osc})/2$ (purple, solid) for $\zeta/\omega_c=0.1$. Here $\delta$ is a quantity that measures how far the last LL $N_v$ is from the lower edge of the gap in $-,e$ band (for an exact definition refer to the text); thus, $\delta=0$ and $\delta=1$ signify the crossing of two consecutive LLs across the gap edge, and corresponds to one period of oscillation. The pattern is repeated giving rise to quantum oscillations.}
\label{fig2}
\end{figure}

\begin{figure}
\centering
\subfigure[]{\includegraphics[width=.45\columnwidth]{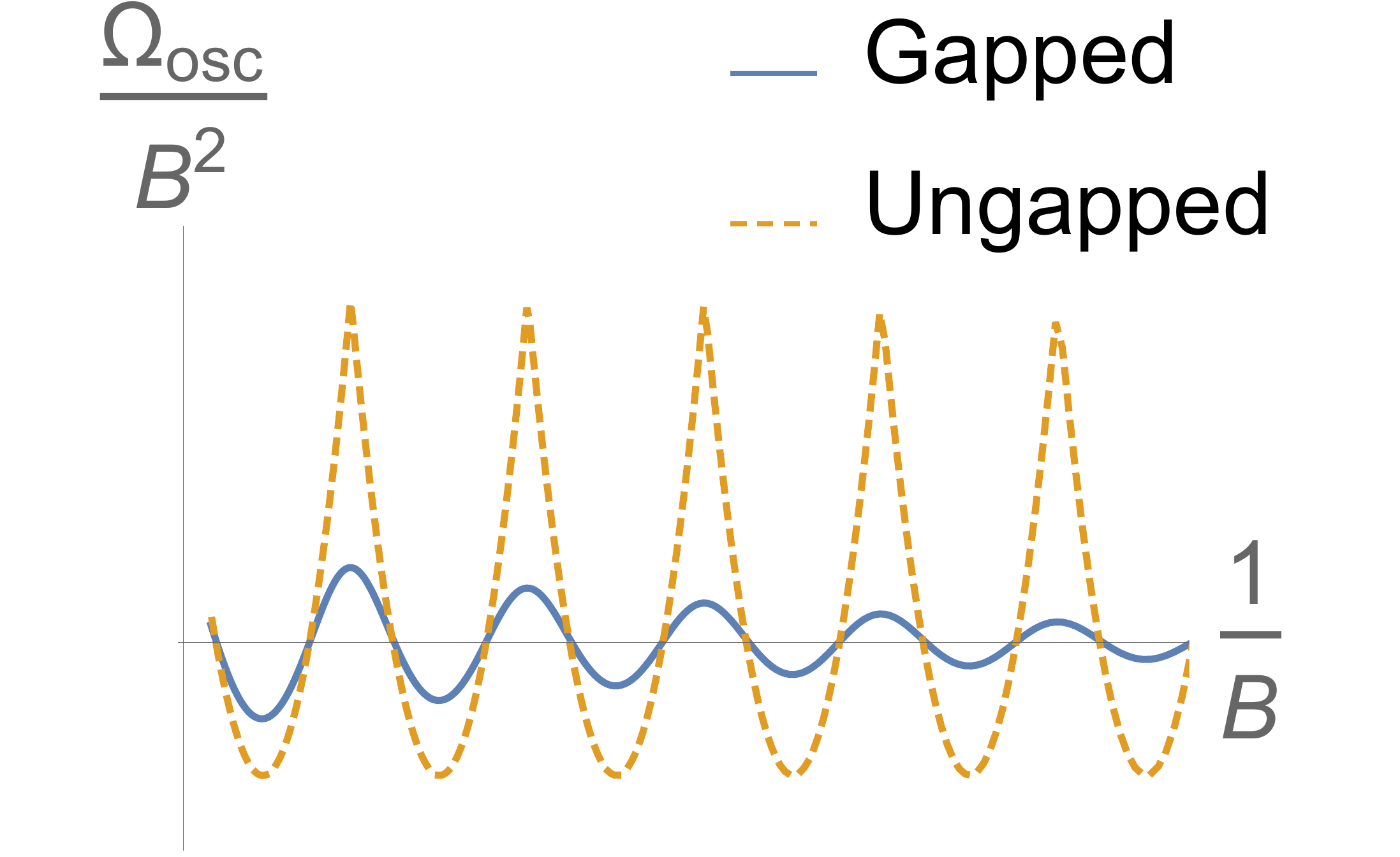}
\label{fig31}}
\quad
\subfigure[]{\includegraphics[width=.45\columnwidth]{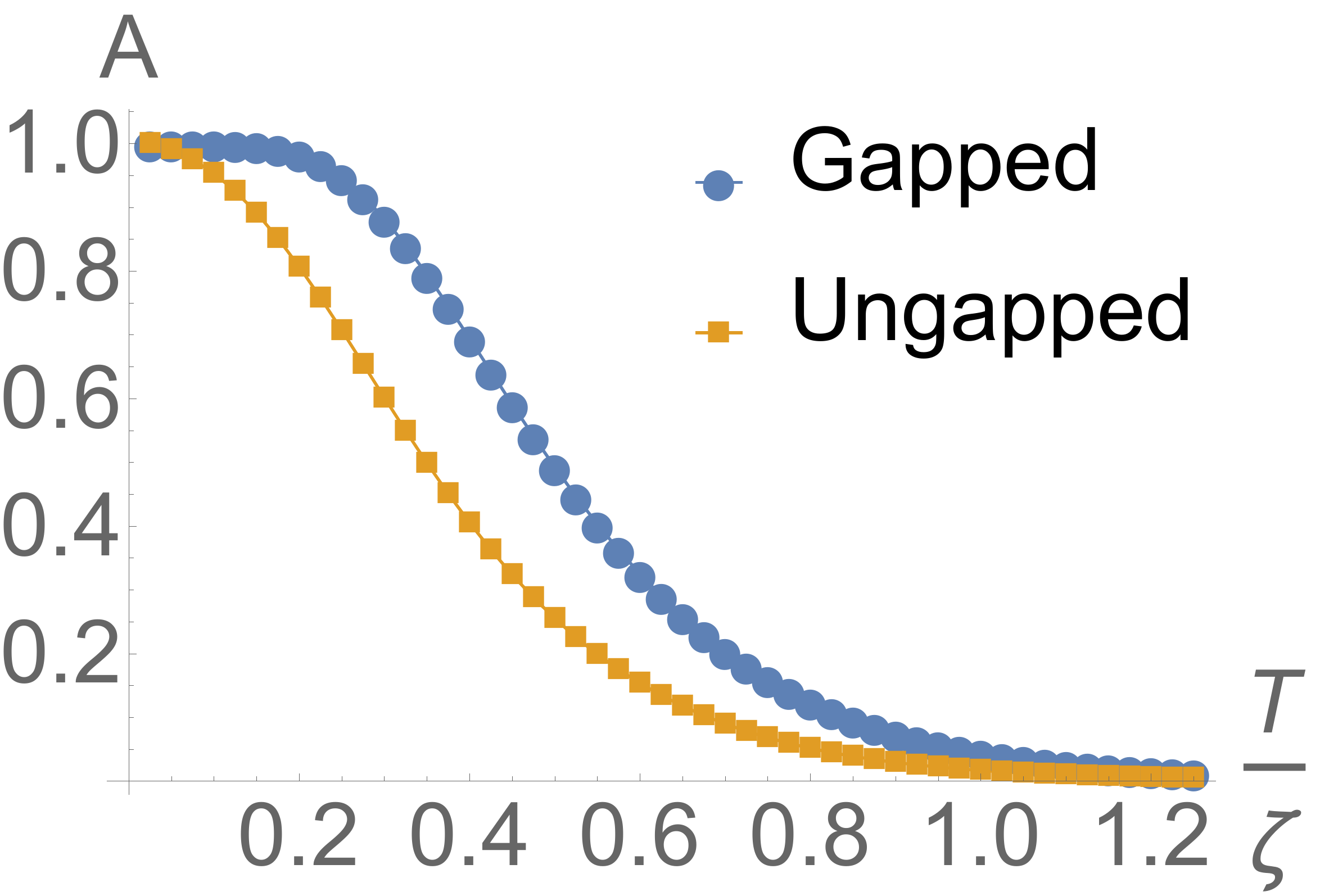}
\label{fig32}}
\subfigure[]{\includegraphics[width=.45\columnwidth]{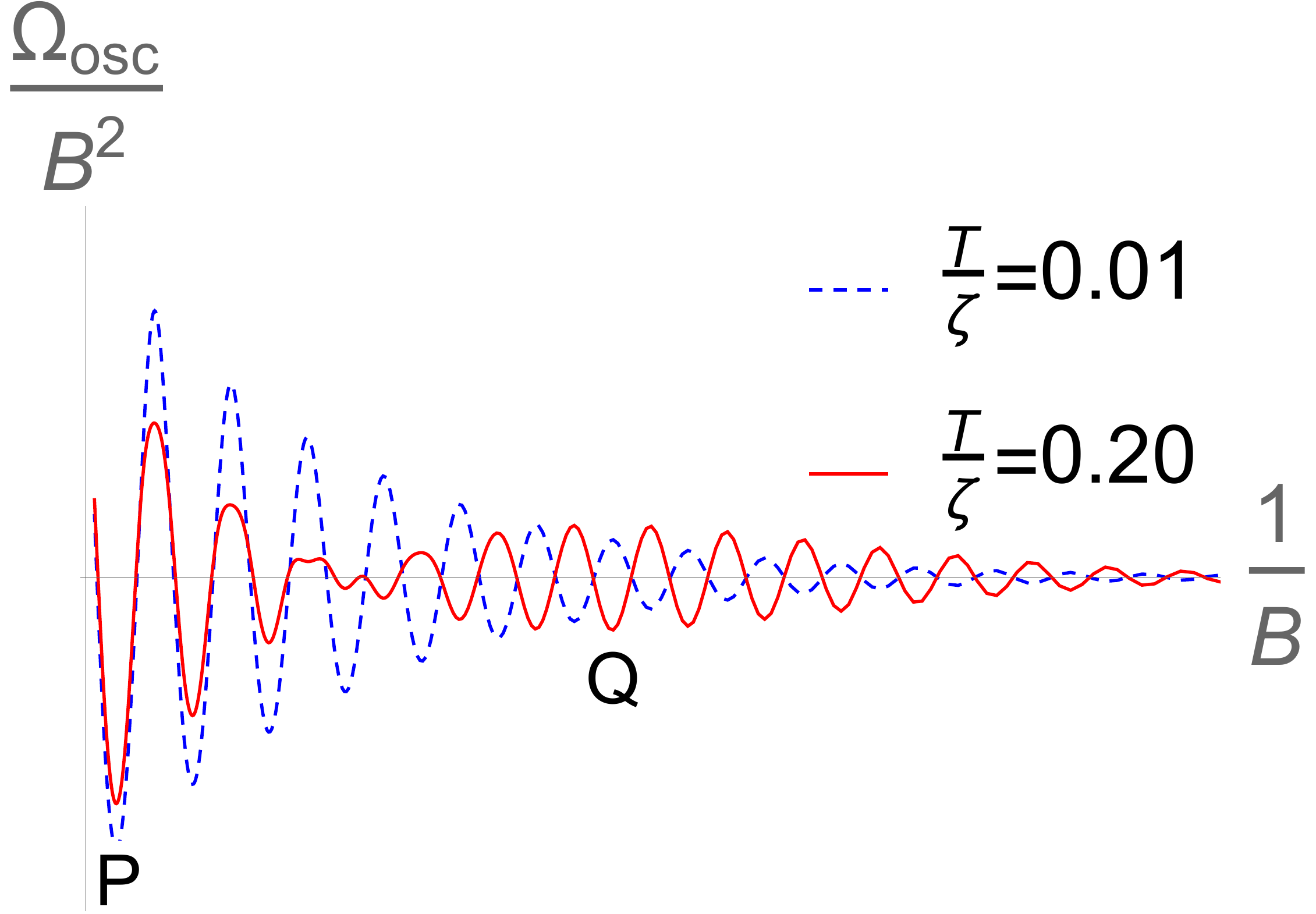}
\label{fig33}}
\quad
\subfigure[]{\includegraphics[width=.45\columnwidth]{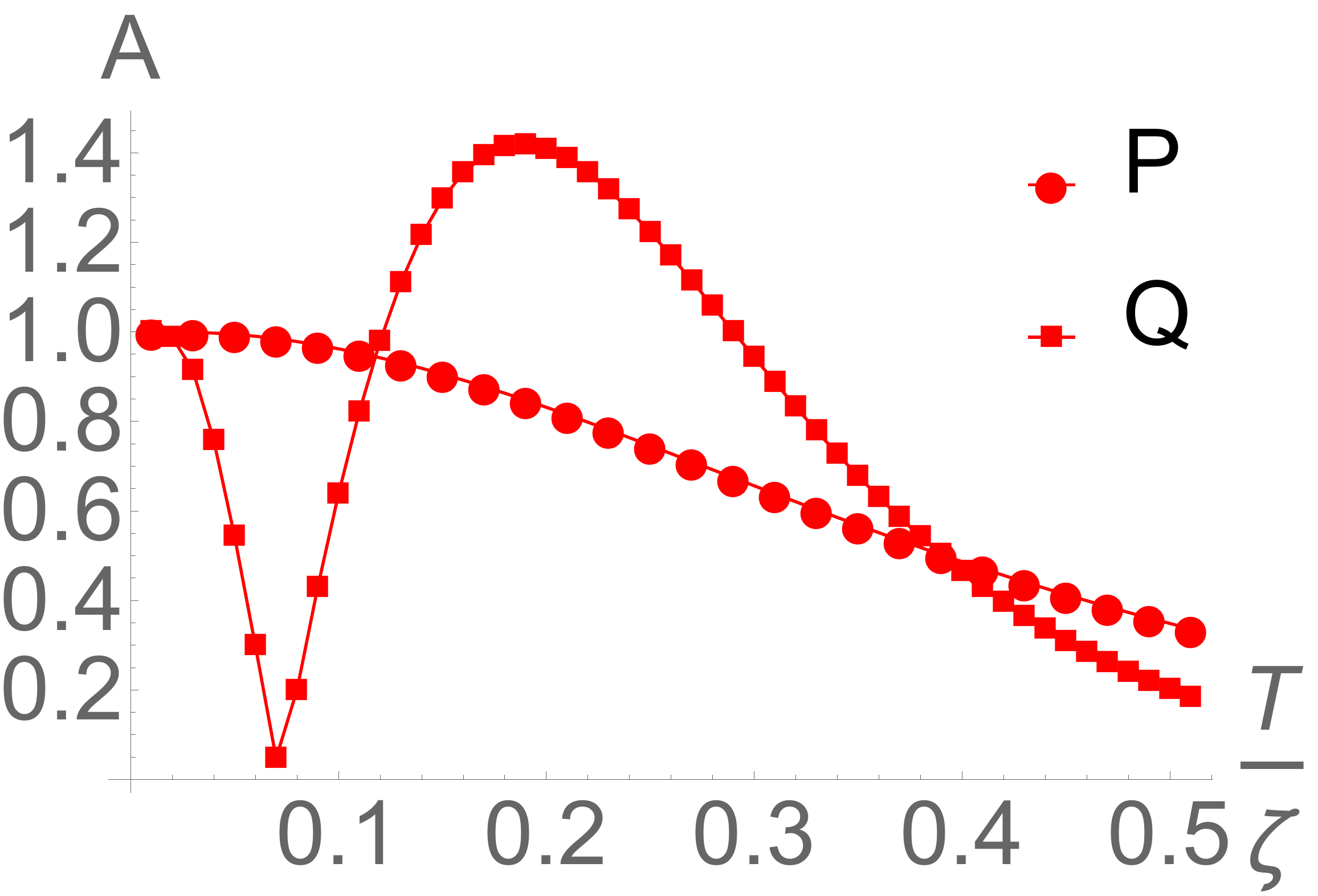}
\label{fig34}}
\caption{Numerical calculations on a lattice model mimicking (\ref{ham}) with two square lattices hybridized to open a gap at quarter filling (see Supplementary Materials \cite{supp} for details). Here $\zeta=0.1t$, where $t$ is the hopping parameter. (a) Oscillations at $T=0$: frequency of oscillations does not change with the introduction of the gap; (b) Dependence of amplitude $A$ on $T$ for $\mu=0$: a plateau appears at $T<\zeta$ in the gapped case (both curves are normalized with their respective $T=0$ values); (c) Oscillations at two different temperatures for $\mu=-0.09t$ (i.e.,$|\mu|/\zeta=0.9$). The beat-like pattern with change of phase by $\pi$ accompanied by an increase in amplitude emerges when temperature is increased. (d) Dependence of oscillations on $T$ at two different field values marked $P$ and $Q$ in (c), normalized with their respective $T=0$ values. }
\label{fig3}
\end{figure}

To compute the discrete sums in Eq.~(\ref{omegaeh}), we use the Euler-MacLaurin formula. This gives for each of the terms $\Omega^{-,e/h}$  a part that varies smoothly with the field and a part that oscillates with the inverse of the field. The behavior is decided by two energy scales: the gap parameter $\zeta$ and the cyclotron frequency of the unhybridized bands, $\omega_c=eB/m$, where $m$ is the cyclotron mass of the unhybridized bands.  In the limits $\zeta/\omega_c\ll 1$ and $\zeta/\omega_c\gg 1$, the oscillating terms $\Omega_{osc}^{-,e/h}$ can be calculated exactly, but they are cumbersome--see Supplementary Materials \cite{supp}. Instead, in Fig.~\ref{fig2}(a) we plot $\Omega_{osc}^{-,e}$ as a function of $\delta$ in the two limits. Here $\delta=(l_B^2/2\pi)[S(E_v)-S(E_{N_v})]$, where $l_B=1/\sqrt{eB}$ is the magnetic length and $S(E)$ is the $\mb{k}$-space area occupied at energy $E$ by an orbit, governing the semiclassical quantization condition $S(E_n)l_B^2=2\pi(n+\gamma)$, with $\gamma$ being a phase \cite{sho}. To understand the meaning of $\delta$, consider the simple case of parabolic bands: $\delta$ reduces to $\Delta/\omega_c-(N_v+1/2)$; thus, it is a measure of how far the last LL is from $E_v$ in the $-,e$ band. Clearly, $0\le\delta<1$, with the limits denoting the crossing of two consecutive LLs across $E_v$, and a plot of $\Omega_{osc}^{-,e}$ vs. $\delta$ gives one period of oscillations--the pattern must be repeated.
Compared to the ungapped case, we find two features as a result of the gap: a reduction in amplitude and a phase offset, with both becoming more pronounced as $\zeta/\omega_c$ increases. Also, in Fig.~\ref{fig2}(b) we compare $\Omega_{osc}^{-,e}$ with $\Omega_{osc}^{-,h}$ along with the total, $\Omega_{osc}=\Omega_{osc}^{-,e}+\Omega_{osc}^{-,h}$. The phases  in $\Omega_{osc}^{-,e}$ and $\Omega_{osc}^{-,h}$ differ by a sign, resulting in a further reduction in amplitude in the total. With this insight, we approximate the curves by their leading Fourier components,
\beq
\frac{\Omega_{osc}^{-,e/h}(B)}{D}\sim \omega_c f\left(\frac{\zeta}{\omega_c}\right)\mathrm{cos}\left[S(E_v)l_B^2-2\pi\gamma\pm\phi\left(\frac{\zeta}{\omega_c}\right)\right],
\label{omegaosceh}
\eeq
with $\pm$ referring to $e/h$  parts, respectively, and $f=1$ and $\phi=0$ at $\zeta=0$. 
Note that the area at the band edge in the gapped case is same as the area at the intersection of the two bands in the ungapped case, i.e., $S(E_v)=S(0)|_{\zeta=0}$. Denoting this area by $S_0$ and adding the $e/h$ contributions in Eq.~(\ref{omegaosceh}), we get
\beq
\frac{\Omega_{osc}(B)}{D}\sim \omega_c f\left(\frac{\zeta}{\omega_c}\right)\mathrm{cos}\left[\phi\left(\frac{\zeta}{\omega_c}\right)\right]\mathrm{cos}\left[S_0l_B^2-2\pi\gamma\right].
\label{omegaosctot}
\eeq
In the limit $\zeta/\omega_c\gg 1$, we find $f(\zeta/\omega_c)\propto1/(\zeta/\omega_c)$ and $\phi(\zeta/\omega_c)\rightarrow\pi/2$. Thus, on opening a gap, the system shows quantum oscillations with frequency equal to that in the ungapped case, but with an amplitude that decays as the product of the functions $f$ and $\mathrm{cos}[\phi]$. This is verified by numerical calculations on a lattice--see Fig.~\ref{fig3}(a). Numerically, we find that the oscillations decay rapidly, becoming inappreciable at $\zeta\gtrsim\omega_c$. 
As an estimate, at field $\sim 10$ T, and mass equal to $0.01$ time the bare electronic mass, gaps $<100$ meV are expected to show oscillations; thus, bands with light masses and narrow gap are required. The main result of the $T=0$ study is the factorization of the amplitude into two terms, and in particular the appearance of the term $\mathrm{cos}[\phi]$. This is unanticipated, and leads to novel effects at non-zero temperature, as shown below. All quantities derived from $\Omega$ by taking appropriate derivatives with respect to the field (magnetization, susceptibility), will inherit oscillations with the same characteristics as well.

\emph{Non-zero temperature.}---We now consider the effects of temperature which is included by averaging $\Omega_{osc}$ at different energies at zero temperature, appropriately weighted by the derivative of the Fermi-Dirac function $f_0$: 
\beq
\Omega_{osc}(\mu,T)=\int_{-\infty}^{\infty}\frac{-\partial f_0(E-\mu)}{\partial E}\Omega_{osc}(E,0)dE.
\label{omegat}
\eeq
Although oscillations at $T=0$ do not depend on the exact value of $\mu$ as long as $\mu$ lies in the gap, the behavior at $T\ne 0$ is dependent on the position of $\mu$ inside the gap (via $f_0$). This extra degree of freedom is a unique feature of insulators not found in metallic systems. 

Consider first the case when $\mu$ is exactly in the middle of the gap, i.e., $\mu=0$. It is important to note that, while $\Omega_{osc}(E,0)$ varies with energy inside the bands, inside the gap it is independent of energy and \emph{nonzero}--it is simply equal to the value at the gap edge, i.e., $\Omega_{osc}^{gap}$ is given by Eq.~(\ref{omegaosctot}). At low temperatures, $T/\zeta\ll 1$, since the integral in Eq.~(\ref{omegat}) gets its dominant contribution from the gap, it implies $\Omega_{osc}$ is independent of $T$ resulting in a plateau--a departure from conventional behavior and supported by numerical calculations [Fig.~\ref{fig3}(b)]. This should be compared to the behavior found in Ref.~\cite{kno} in the same regime: instead of a plateau the dependence was found to be non-monotonic with a maximum. This implies that the behavior found in Ref.~\cite{kno} is a result of the extreme particle-hole asymmetry arising from the flat band in their model and is not a generic feature of an insulator. In the other limit, $T/\zeta\gg 1$, the dominant contribution to the integral comes from the two bands and the gap can be neglected. One then recovers the exponential decay due to dephasing typical of metals, provided by LK formalism.

Next, we consider $\mu\ne 0$ with $|\mu|<\zeta$, i.e., the system remains an insulator but $\mu$ lies asymmetrically in the gap.
Such a situation can arise due to impurities (extrinsic semiconductors) or can be imposed by an external gate in experiments.
The behavior for $T\ll\zeta-|\mu|$ and $T\gg\zeta$ is similar to that in the case $\mu=0$. But, in the intermediate regime, $\zeta-|\mu|\lesssim T\lesssim \zeta$, the new scale $|\mu|$ introduces surprising new features. The effect of temperature is no longer restricted to the overall amplitude. Instead, as seen in Fig.~\ref{fig3}(c) obtained numerically, oscillations go to zero at a critical value of the inverse field, $1/\omega_c^{\ast}$, and then change their phase by $\pi$ and grow again. With increase in temperature, $1/\omega_c^{\ast}$ moves to the left with increase in amplitude to its right and decrease in amplitude to its left [Fig.~\ref{fig3}(d)]. In the limit of strong asymmetry in $\mu$, i.e, $\zeta-|\mu|\ll\zeta$, such a behavior can be explained analytically. When $\mu=0$, oscillations arising from $+E$ and $-E$ get equal thermal weight while computing the average in Eq.~(\ref{omegat}).
This particle-hole symmetry in the averaging is lost when $\mu\ne 0$, even though the bands are still particle-hole symmetric. 
One can show that (see Supplementary Materials \cite{supp}) this leads to an extra phase in Eq.~(\ref{omegaosceh}) on top of the overall prefactor: the zero temperature result  modifies into $\Omega_{osc}^{e/h}(T)/D\sim\omega_cA(T)f(\zeta/\omega_c)\mathrm{cos}[S_0l_B^2-2\pi\gamma\pm\phi(\zeta/\omega_c)\pm\psi(T)]$. 
Adding the two contributions, we have
\begin{eqnarray}
\frac{\Omega_{osc}(T)}{D}&\sim&\omega_cA\left(\frac{T}{\omega_c},\frac{T}{\zeta},\frac{|\mu|}{\zeta}\right)f\left(\frac{\zeta}{\omega_c}\right)\mathrm{cos}\left[S_0l_B^2-2\pi\gamma\right]\nonumber\\
&\times&\mathrm{cos}\left[\phi\left(\frac{\zeta}{\omega_c}\right)+\psi\left(\frac{T}{\omega_c},\frac{T}{\zeta},\frac{|\mu|}{\zeta}\right)\right].
\label{eq:mut}
\end{eqnarray}
Whereas $A(T)$ is a decreasing function as in the case of $\mu=0$, $\psi(T)$ increases with $T$ and reaches a maximum determined by $|\mu|$, before going to zero. Eq.~(\ref{eq:mut}) leads to an unusual beat-like (but not exactly a beat) pattern driven by temperature that matches Fig.~\ref{fig3}(c).  This arises from a competition between $\phi$ and $\psi$ in the second cosine function in Eq.~(\ref{eq:mut}). Recall, $\phi\rightarrow\pi/2$ as $1/\omega_c\rightarrow\infty$. In the presence of $\psi$, $\phi+\psi$ reaches $\pi/2$ at a finite value of $1/\omega_c$. This is where oscillations go to zero and is the origin of the critical field. The sign of the cosine function is opposite on either side of $1/\omega_c^{\ast}$ leading to the phase change by $\pi$. Note the crucial role $\phi$, arising in the $T=0$ theory, plays here: it forces the effect of $T\ne 0$ to be no longer separable from the $T=0$ contribution, and is at the heart of the unusual behavior.
The $\pi$ phase shift here is  temperature driven and not topological in origin, unlike in metals where it originates due to Berry phase \cite{zha,ber}. Also, without proper insight, the pattern could easily be misconstrued as resulting from two Fermi surface pockets in metallic systems.

\begin{figure}
\centering
\subfigure[]{\includegraphics[width=.46\columnwidth]{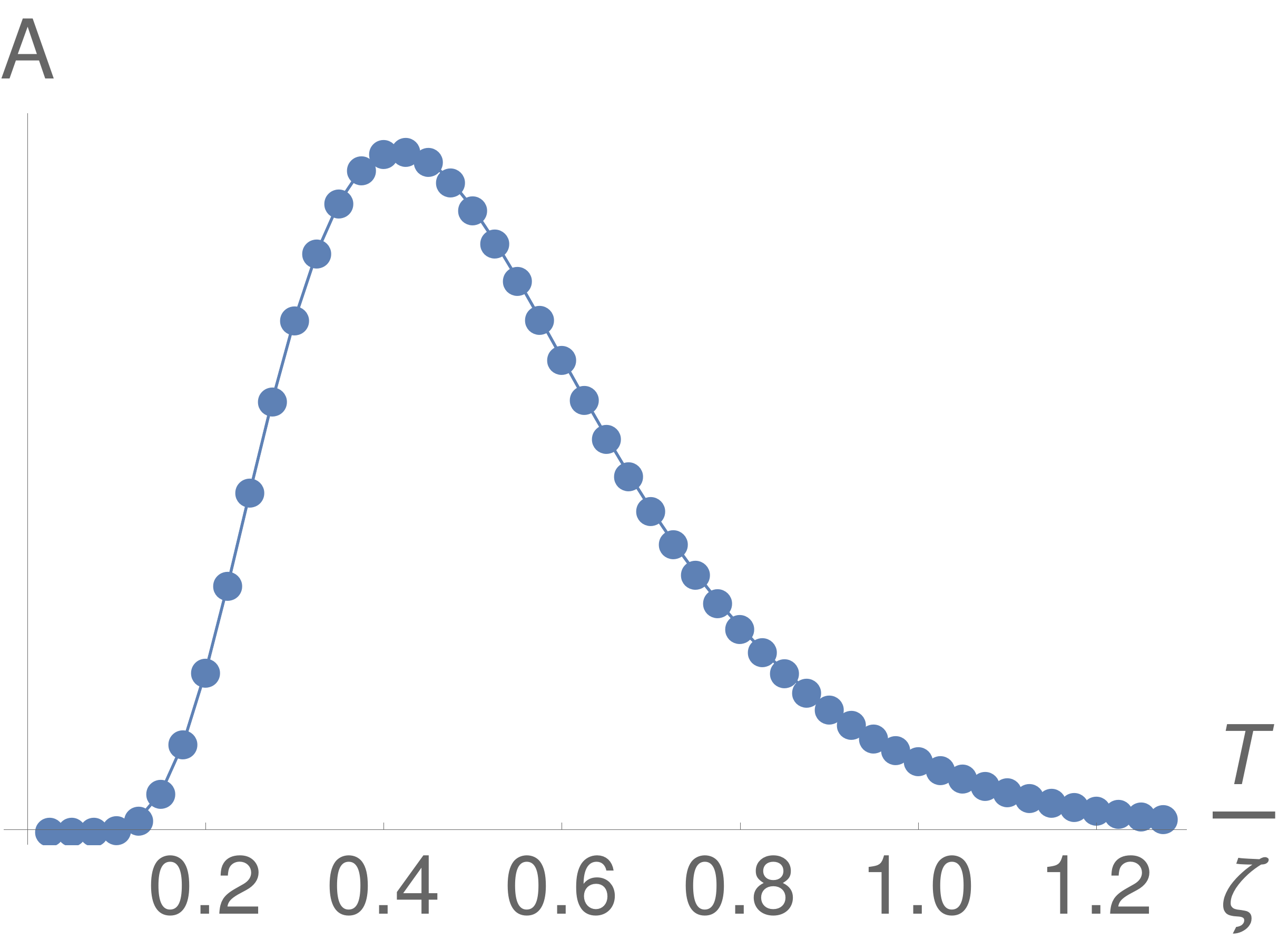}
\label{fig41}}
\quad
\subfigure[]{\includegraphics[width=.46\columnwidth]{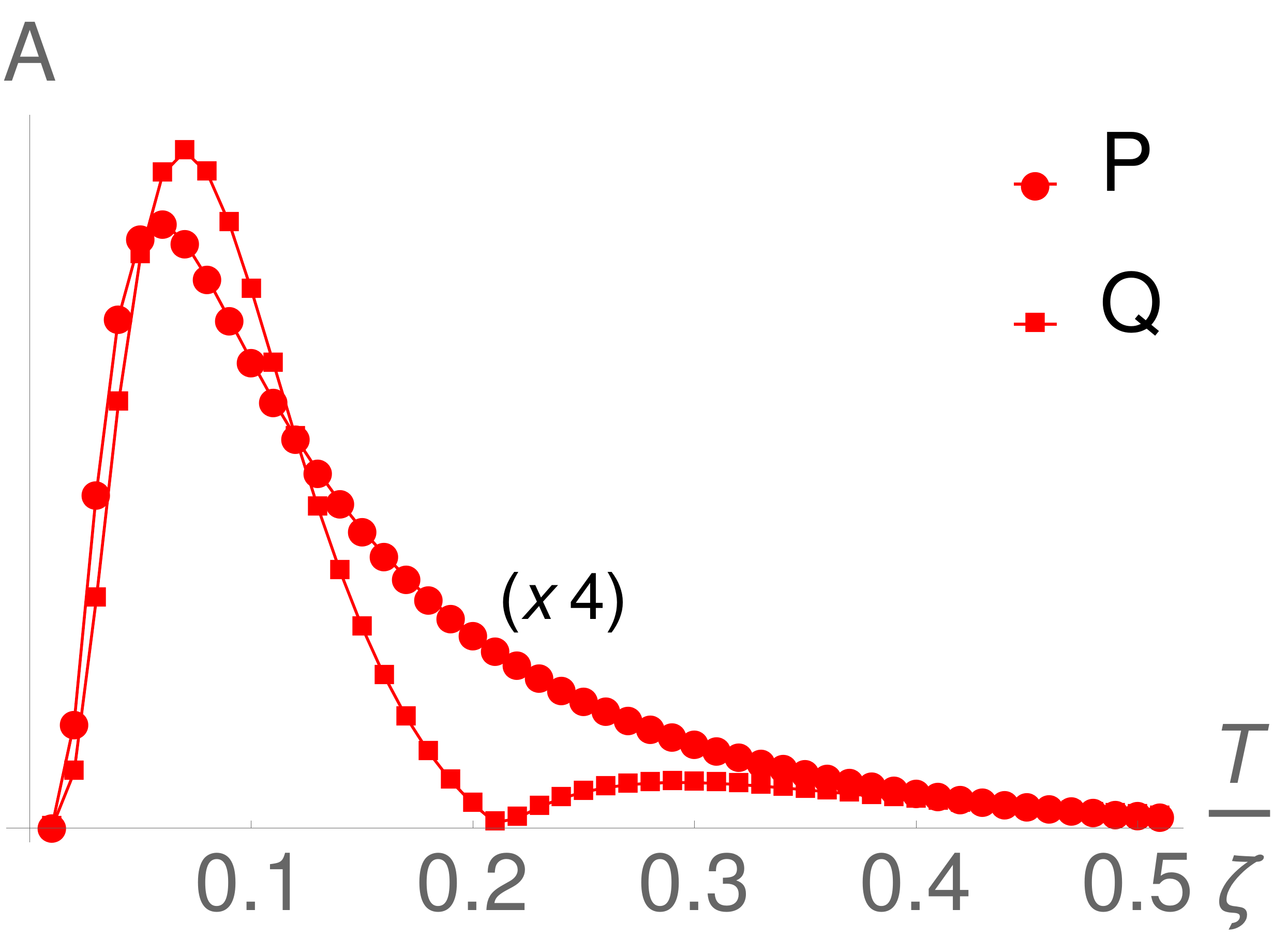}
\label{fig42}}
\caption{Amplitude  of oscillations $A$ (arbitrary units) vs. $T$  for density of states in the gapped case obtained from numerical calculations on a lattice (same as in Fig.~\ref{fig3}) with $\zeta=0.1t$ for (a) $\mu=0$ and (b) $\mu=-0.09t$. The figures should be compared to Figs.~\ref{fig3}(b) and (d), respectively. The behavior for $T\ll\zeta$ is clearly different. Note, however, the extra non-trivial features in the case $\mu\ne 0$ found in Fig.~\ref{fig3}(d) persist in Fig.~\ref{fig4}(b) as well.}
\label{fig4}
\end{figure}

\emph{Grand canonical potential vs. density of states.}---It is usually thought that quantum oscillations in all physical observables have the same temperature dependence \cite{sho,mar}.
In an insulator, however, this is no longer true. Unlike quantities which are derived from the grand canonical potential, such as magnetization and susceptibility, quantities which depend on the DOS, such as resistivity and quantum capacitance, will obviously vanish in the gap at $T=0$. This implies that the averaging in Eq.~(\ref{omegat}) at $T\ne 0$ is different: it is still governed by an equation similar to Eq.~(\ref{omegat}) (with $\Omega$ replaced by DOS $\rho$), except that now the integral gets no contribution from the gap. This results in a temperature dependence that is non-monotonic, arising from a competition between thermal activation and dephasing.
Numerical calculations on the lattice also confirm this behavior--see Fig. \ref{fig4}. Note, however, the non-trivial features for $\mu\ne 0$ survive: oscillations show a similar pattern as in Fig.~\ref{fig3}(c) changing their phase by $\pi$ at a critical field with unusual temperature dependence as shown in Fig.~\ref{fig4}(b).

\begin{figure}
\includegraphics[width=0.7\columnwidth]{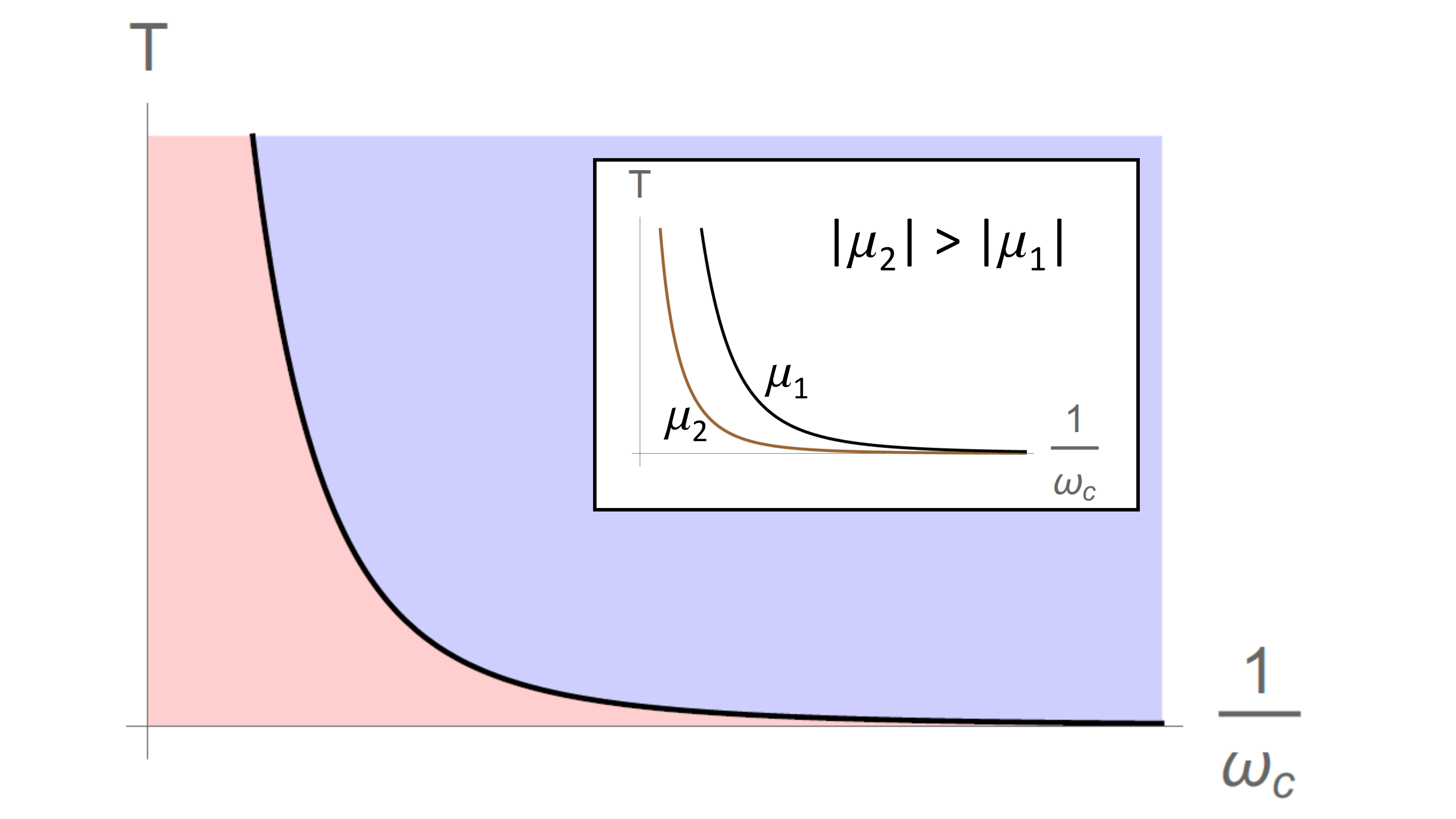}
\caption{The two shaded regions differ by a phase $\pi$ in oscillations. The separatrix is a schematic variation of the critical field with temperature. \emph{Inset}: The separatrix moves to the left as $|\mu|$ increases. Numerical calculations on a lattice confirm this behavior--see Supplementary Materials \cite{supp}.}
\label{fig5}
\end{figure}

It is a remarkable coincidence that the non-tirival features in the case of $\mu\ne 0$ discussed above also arise from a different physical mechanism, viz., non-trivial topology in topological insulators \cite{zha}. The question then arises, how to distinguish in an experiment which physical mechanism is at play. We point out two key differences between the two scenarios: first, in a topological insulator the gap is a function of the field, and the critical field marks the point when the Landau levels overlap in the gap making the system metallic. Thus, at $T=0$, the DOS is zero on the insulating side but non-zero on the metallic side. In the case considered here, the system stays gapped at all fields; therefore, the DOS is zero at $T=0$ on both sides of the critical field. Consequently, while both curves in Fig.~\ref{fig4}(b) in our case start from zero at $T=0$, one of the curves start from a non-zero value at $T=0$ in the case of topological insulators [cf. Figs. 4(c) and 5(c) in \cite{zha}]. Second--and more importantly--the critical field in a topological insulator is a function of the band parameters only, and is independent of temperature. In our case, it is a function of both $T$ and $|\mu|$. Thus, with increase in temperature, not only will oscillations behave differently on either sides of the critical field, the critical point itself will move to the left on $1/\omega_c$ axis--see Fig.~\ref{fig5}. This should be compared with Figs. 4(d) and 5(d) in Ref.~\cite{zha}. 

It is important to note that not all types of insulators will show quantum oscillations. If we change the sign of  $\Delta$ in Eq.~(\ref{ham}) so that there is no band inversion, it is obvious that such a model will show no oscillations. Thus, in addition to having a narrow gap, band inversion that leads to a closed loop at the band edge is needed.
Although, our results are derived for a 2D system, they apply as well to 3D systems, as long as the system is described by Eq.~(\ref{ham}): as in metals oscillations will arise from extremal orbits, but with properties modified due to the gap in a manner described above. Quantum oscillations, traditionally used to study metallic systems, could soon become a useful experimental tool to study narrow-gap systems with inverted bands.

\begin{acknowledgements}
This work was supported by the ANR DIRACFORMAG project, grant ANR-14-CE32-0003 of the French Agence Nationale de la Recherche.
\end{acknowledgements}

\begin{widetext}

\section{Supplementary Materials}

\section{Oscillations at zero temperature}

We consider two overlapping bands, $\varepsilon_{\mb{k}}-\Delta$ and $\Delta-\varepsilon_{\mb{k}}$, with an overlap of $2\Delta$, hybridized by a parameter $\zeta$ to open a gap of $2\zeta$. The grand canonical potential $\Omega$ in the presence of a magnetic field $B$ at $T=0$ for such a system reads  $\Omega=D\sum_{E_n^{-}\le E_v}[E_n^{-}-E_v]$,
where 
\beq
E_n^{-}=-\sqrt{(\varepsilon_n-\Delta)^2+\zeta^2}
\label{landaulevel}
\eeq
denotes the lower occupied band ($\varepsilon_n$ denotes the Landau levels (LLs) corresponding to $\ve_{\mb{k}}$ in the absence of the gap), $E_v=-\zeta$ is the lower edge of the gap, and $D=geB/2\pi$ is the degeneracy factor (g denotes any extra degeneracy from internal degrees of freedom).
One can decompose the sum into two parts:
\beq
\frac{\Omega(B)}{D}=\sum_{n=0}^{N_v}[E_n^{-,e}-E_v]+\sum_{N_v+1}^{\Lambda}[E_n^{-,h}-E_v]=\frac{\Omega^{-,e}}{D}+\frac{\Omega^{-,h}}{D},
\label{omegaehsup}
\eeq
where $E_n^{-,e/h}$ denote the electron-like and hole-like parts of the lower filled band, respectively, $N_v$ is the highest LL with energy smaller than $E_v$ in the electron-like part of the spectrum, and $\Lambda$ is a cutoff for the hole-like part of the band ($E^{-}_{\Lambda}\sim$ bandwidth). To find an expression for the oscillating part of the grand canonical potential we need to carry out the discrete summations in Eq.~(\ref{omegaehsup}). We first concentrate on $\Omega^{-,e}$. There are two ways to carry out the discrete sum: Poisson summation formula and Euler-Maclaurin formula. Here, we follow the latter approach since it brings out the physics more clearly. 
Euler-Maclaurin formula for a discrete sum of any function $f(r)$ over $r$ reads
\beq
\sum_{r=0}^Rf(r)=\int_0^Rf(r)dr+\frac{1}{2}[f(R)+f(0)]+\frac{1}{12}[f'(R)-f'(0)]+\cdots,
\label{eulermac}
\eeq
where $r$ on the right hand side is treated as a continuous variable. The formula can be thought of as an extension of the trapezoidal rule to approximate the integral of a function in some interval by a sum over discrete values of the function within that interval, with corrections from the end point of the interval. The corrections are small if the number of discrete points summed over is large. Alternatively, this implies that the function must vary smoothly between any two points in the interval, $r$ and $r+1$, for all $r$, i.e., there is no scale in the problem that is smaller than the difference between any two discrete points. In metals, since $\omega_c\ll \mu$ is the smallest energy scale in the problem, summation over the LLs can be well approximated by Eq.~(\ref{eulermac}). In the gapped case, however, the presence of an extra scale $\zeta$ means we cannot apply Eq.~(\ref{eulermac}) directly. Indeed, in the limit $\zeta/\omega_c\ll 1$, $\zeta$ is the smallest energy scale in the problem, and not $\omega_c$. One way to circumvent the problem is to note that in this limit, only the last LL below the Fermi level is effectively affected by $\zeta$. Hence, we first separate the last LL from the summation:
\beq
\frac{\Omega^{-,e}}{D}=\sum_0^{N_v-1}[E^-_n-E_v]+[E^-_{N_v}-E_v].
\label{split}
\eeq
With this separation, we can now perform the sum in the first term using Eq.~(\ref{eulermac}) (we drop the superscripts $-,e$ henceforth for brevity): 
\beq
\sum_0^{N_v-1}[E_n-E_v]\approx\int_0^{N_v-1}\left[E(n)-E_v\right]dn+\frac{1}{2}\left[\{E(N_v-1)-E_v\}+\{E(0)-E_v\}\right]+\frac{1}{12}\left[E'(N_v-1)-E'(0)\right].
\label{omegasum}
\eeq
The general relation between $E$ and $n$ is given by the semiclassical quantization condition,
\beq
S(E)l_B^2=2\pi(n+\gamma),
\label{quantcon}
\eeq
where $S(E)$ gives the area in reciprocal space as a function of $E$, $l_B$ is the magnetic length, $n$ is the LL index, and $\gamma$ is a phase. To proceed further, it is useful to define a variable $x$ in place of $n+\gamma$ and rewrite the quantization condition as
\beq
S(E)l_B^2=2\pi x.
\eeq
Let $X$ be the value $x$ takes at the band edge, i.e., $S(E_v)l_B^2=2\pi X$. With change in magnetic field, the LLs move, and each time a LL crosses $E_v$, $X$ changes by one. With this in mind define $\delta=X-(N_v+\gamma)$ so that $0\le\delta\le1$. Inserting this in Eq.~(\ref{omegasum}), we have
\beq
\sum_0^{N_v-1}[E_n-E_v]\approx\int_{\gamma}^{X-\delta-1}[E(x)-E_v]dx+\frac{1}{2}[\{E(X-\delta-1)-E_v\}+\{E(\gamma)-E_v\}]+\frac{1}{12}[E'(X-\delta-1)-E'(\gamma)].
\label{omegasumX}
\eeq
Clearly, the terms that depend on $\delta$ in the above expression are responsible for oscillations (the remaining terms give rise to a field dependent continuous background). Splitting the integral in Eq.~(\ref{omegasumX}) as $\int_{\gamma}^{X-\delta-1}\rightarrow\int_{\gamma}^{X}+\int_{X}^{X-\delta-1}$, followed by a change of variable, and collecting all the terms that depend on $\delta$, we have
\beq
\sum_0^{N_v-1}[E_n-E_v]_{osc}=\int_{\delta+1}^{0}[E(X-x)-E_v]dx+\frac{1}{2}[E(X-\delta-1)-E_v]+\frac{1}{12}[E'(X-\delta-1)].
\label{oscsum}
\eeq
Near the edge, using Eq.~(\ref{landaulevel}), one can write
\beq
E(X+\delta)=-\sqrt{\omega_c^2\delta^2+\zeta^2},
\label{edgespectrum}
\eeq
where $\omega_c=\partial\ve/\partial n$. In arriving at Eq.~(\ref{edgespectrum}), we have linearized the original unhybridized band near the edge. This is justified as long as the effective mass of the unhybridized band does not change on the scale of $\zeta$, which is a reasonable assumption. In this sense, our results are universal and do not depend on the details of the spectrum. Inserting Eq.~(\ref{oscsum}) in Eq.~(\ref{split}), and making use of Eq.~(\ref{edgespectrum}), we finally derive an expression for $\Omega^{-,e}_{osc}$ in terms of $\delta$:
\begin{equation}
\frac{\Omega^{-,e}_{osc}}{D}=\frac{\omega_c}{2} \left[a (1-2 \delta )-2 \sqrt{a^2+\delta ^2}+\frac{\delta  \left(6 a^2+6 (\delta +1)^2+1\right)+1}{6 \sqrt{a^2+(\delta +1)^2}}-a^2\log \left(\frac{a}{\sqrt{a^2+(\delta +1)^2}+\delta +1}\right)\right],
\label{deltasmall}
\end{equation}
where $a=\zeta/\omega_c$.
In the other limit $\zeta/\omega_c\gg 1$, $\omega_c$ is the smallest scale in the problem, and, therefore, one can carry out the  usual Euler-Maclaurin expansion as in Eq.~(\ref{eulermac}). Going through similar steps, we have
\beq
\frac{\Omega^{-,e}_{osc}}{D}=\frac{\omega_c}{2}\frac{1}{a}\left[\frac{\delta^3}{3}-\frac{\delta^2}{2}+\frac{\delta}{6}\right].
\label{deltalarge}
\eeq
Also, from symmetry it can be shown that
\beq
\Omega_{osc}^{-,h}(\delta)=\Omega_{osc}^{-,e}(1-\delta).
\label{hole}
\eeq
The above expressions may be compared with the case of a metal (ungapped case) \cite{shoenberg}:
\beq
\frac{\Omega_{osc}}{D}=\frac{\omega_c}{2 }\left[\delta^2-\delta+\frac{1}{6}\right]
\label{zeta0}
\eeq
The above expressions are valid only for $0\le\delta\le 1$; on reaching the boundary of this limit, $\delta$ restarts from zero, and the pattern repeats itself, rendering periodic oscillations in the grand canonical potential. Plots of Eqs.~(\ref{deltasmall}), (\ref{deltalarge}), (\ref{hole}), and (\ref{zeta0}), along with Eq.~(\ref{omegaehsup}) appear in Fig. 2 in the main text. Approximations of  expressions (\ref{deltasmall}) and (\ref{hole}) in terms of their leading Fourier components appear in Eq. (4) in the main text.

\section{Effect of temperature when $\mu\ne 0$}

Interesting new features arise in quantum oscillations that are not found in the metallic case when temperature is non-zero, particularly in the case when the chemical potential lies asymmetrically in the gap, i.e., $\mu\ne 0$, with $|\mu|<\zeta$. This happens in an intermediate regime, $\zeta-|\mu|\lesssim T\lesssim\zeta$. Here, we derive the dependence of the grand canonical potential on temperature in this case.

The effect of temperature on quantum oscillations is given by:
\beq
\Omega_{osc}(\mu,T)=\int_{-\infty}^{\infty}\frac{-\partial f_0(E-\mu)}{\partial E}\Omega_{osc}(E,0)dE.
\eeq
We rewrite it as
\beq
\Omega_{osc}(\mu,T)=\sum_{\alpha=e/h}\int_{-\infty}^{-\zeta}\frac{\Omega_{osc}^{-,\alpha}(E,0)}{1+\mathrm{cosh}\left[\frac{E-\mu}{T}\right]}\frac{dE}{2T}+\int_{-\zeta}^{\zeta}\frac{\Omega_{osc}^{gap,\alpha}(E,0)}{1+\mathrm{cosh}\left[\frac{E-\mu}{T}\right]}\frac{dE}{2T}+\int_{\zeta}^{\infty}\frac{\Omega_{osc}^{+,\alpha}(E,0)}{1+\mathrm{cosh}\left[\frac{E-\mu}{T}\right]}\frac{dE}{2T},
\label{omegatt}
\eeq
where the contributions from from the lower band, gap, and upper band, along with the electron-like and hole-like contributions within each, have been explicitly shown. In the limit of strong asymmetry in the position of $\mu$, i.e., $\zeta-|\mu|\ll\zeta$, because we are interested in the intermediate regime of temperatures, $\zeta-|\mu|\lesssim T\lesssim\zeta$, we can forget the contribution from one of the bands. Without any loss of generality, we assume $\mu<0$ (by symmetry the results apply to $\mu>0$ as well). In this case, we forget the contribution from the upper band, and rewrite Eq.~(\ref{omegatt}) as
\beq
\Omega_{osc}(\mu,T)\approx\sum_{\alpha=e/h}\int_{-\infty}^{-\zeta}\frac{\Omega_{osc}^{-,\alpha}(E,0)}{1+\mathrm{cosh}\left[\frac{E-\mu}{T}\right]}\frac{dE}{2T}+\int_{-\zeta}^{\infty}\frac{\Omega_{osc}^{gap,\alpha}(E,0)}{1+\mathrm{cosh}\left[\frac{E-\mu}{T}\right]}\frac{dE}{2T}.
\label{temp1}
\eeq
Note that $\Omega_{osc}^{gap}$ does not vary inside the gap, and is equal to its value at the edge of the gap, i.e., $\Omega_{osc}^{gap,e/h}=\Omega_{osc}^{e/h}(-\zeta)$ given by Eq. (4) in the main text:
\beq
\frac{\Omega_{osc}^{e/h}(-\zeta)}{D}\sim \omega_c f\left(\frac{\zeta}{\omega_c}\right)\mathrm{cos}\left[S_0l_B^2-2\pi\gamma\pm\phi\left(\frac{\zeta}{\omega_c}\right)\right],
\label{omegaoscehsup}
\eeq
where $S_0=S(-\zeta)$, i.e., $S_0$ is the area at the band edge in the gapped case, which is same as the area at the intersection of the two bands in the ungapped case. 
The form of $\Omega_{osc}^{-,e/h}(E)$ can be obtained by generalizing Eq.~(\ref{omegaoscehsup}): $\Omega_{osc}^{-,e/h}(E)/D\omega_c\sim f^{-,e/h}(E)\mathrm{cos}[S^{-,e/h}(E)l_B^2-2\pi\gamma\pm\phi^{-,e/h}(E)]$. To proceed further, we need to find the dependence of the functions $f$, $S$,  and $\phi$ on $E$. First, we note that, while $S$ varies with $E$ even in the absence of gap and $\zeta$ provides additional correction, the functions $f$ and $\phi$ are entirely due to the gap, and are absent in the ungapped case. Therefore, in the limit $\zeta/\omega_c<1$--which is when the oscillations are expected to be appreciable--to leading order one can neglect the energy dependence of $f$ and $\phi$ and simply consider the energy dependence of $S$. 
Next, we note that at the edge, i.e., at $E=-\zeta$, $E(n)$ is maximum. This implies $\partial E/\partial n=\partial E/\partial S=0$ and $\partial^2 E/\partial S^2<0$ at $E=-\zeta$, where we have used the fact that $S\propto n$ through the semiclassical quantization condition (\ref{quantcon}). Near the edge, then, one can write \cite{comment1} $S^{-,e/h}(E)=S(-\zeta)+\mathrm{sgn}(e/h)\sqrt{\frac{E+\zeta}{E''(-\zeta)/2}}$, where $\mathrm{sgn}(e/h)=\mp 1$ and $E''(-\zeta)=\partial^2 E/\partial S^2|_{S(-\zeta)}$.
Using complex notation for notational convenience, Eq.~(\ref{temp1}) reduces to
\begin{equation}
\frac{\Omega_{osc}(\mu,T)}{D}\sim \omega_c\mathrm{Re}[fe^{i(S_0l_B^2-2\pi\gamma)}(\mathcal{I}^{e}+\mathcal{I}^{h})],
\label{temptot}
\end{equation}
where
\begin{eqnarray}
\mathcal{I}^{e}&=&e^{i\phi}\left\{\int_{-\infty}^{-\zeta}\frac{e^{-i\alpha}}{1+\mathrm{cosh[\frac{E-\mu}{T}]}}\frac{dE}{2T}+\int_{-\zeta}^{\infty}\frac{1}{1+\mathrm{cosh[\frac{E-\mu}{T}]}}\frac{dE}{2T}\right\},
\nonumber\\
\mathcal{I}^{h}&=&e^{-i\phi}\left\{\int_{-\infty}^{-\zeta}\frac{e^{i\alpha}}{1+\mathrm{cosh[\frac{E-\mu}{T}]}}\frac{dE}{2T}+\int_{-\zeta}^{\infty}\frac{1}{1+\mathrm{cosh[\frac{E-\mu}{T}]}}\frac{dE}{2T}\right\},
\label{integralseh}
\end{eqnarray}
with
\beq
\alpha=l_B^2\sqrt{\frac{E+\zeta}{E ''(-\zeta)/2}}.
\eeq
Integrals $\mathcal{I}^{e/h}$ are, in general, complex, and conjugates of each other, i.e., $\mathcal{I}^{e}=\mathcal{I}^{h\ast}$. 
Writing $\mathcal{I}^{e}=e^{i\phi}\{Ae^{i\psi}\}$, and inserting it into Eq.~(\ref{temptot}), we find
\begin{equation}
\frac{\Omega_{osc}(T)}{D}\sim\omega_cA\left(\frac{T}{\omega_c},\frac{T}{\zeta},\frac{|\mu|}{\zeta}\right)f\left(\frac{\zeta}{\omega_c}\right)\mathrm{cos}\left[S_0l_B^2-2\pi\gamma\right]
\mathrm{cos}\left[\phi\left(\frac{\zeta}{\omega_c}\right)+\psi\left(\frac{T}{\omega_c},\frac{T}{\zeta},\frac{|\mu|}{\zeta}\right)\right],
\label{delmunotzero}
\end{equation}
where we have explicitly stated the dependence of different functions on the different parameters (made dimensionless appropriately). This result appears in Eq. (7) in the main text.

 \begin{figure}
\centering
\subfigure[]{\includegraphics[scale=0.31,trim={3.5cm 0 3.5cm 0},clip]{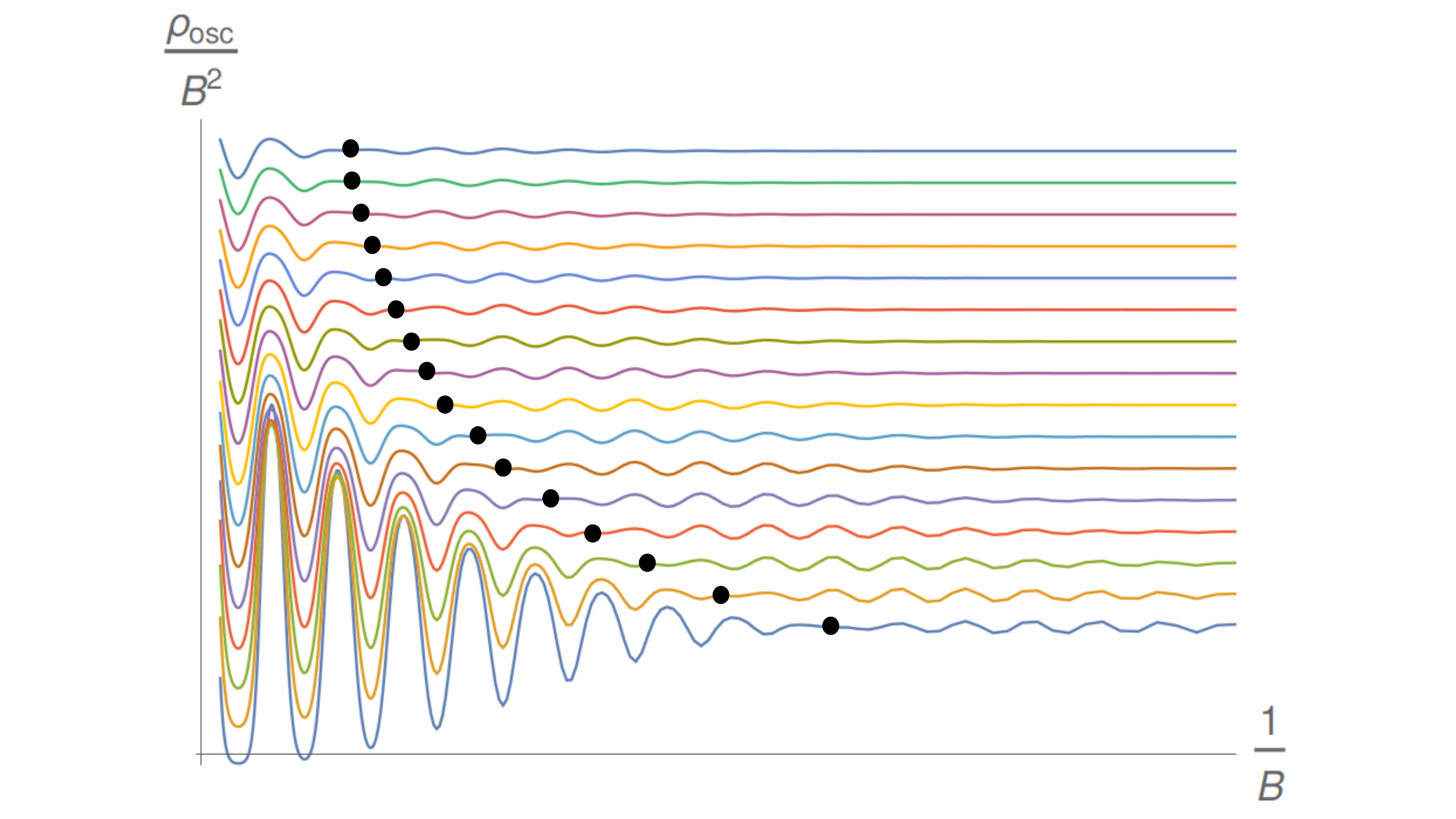}
\label{fig11_supp}}
\quad
\subfigure[]{\includegraphics[scale=0.31,trim={3.5cm 0 3.5cm 0},clip]{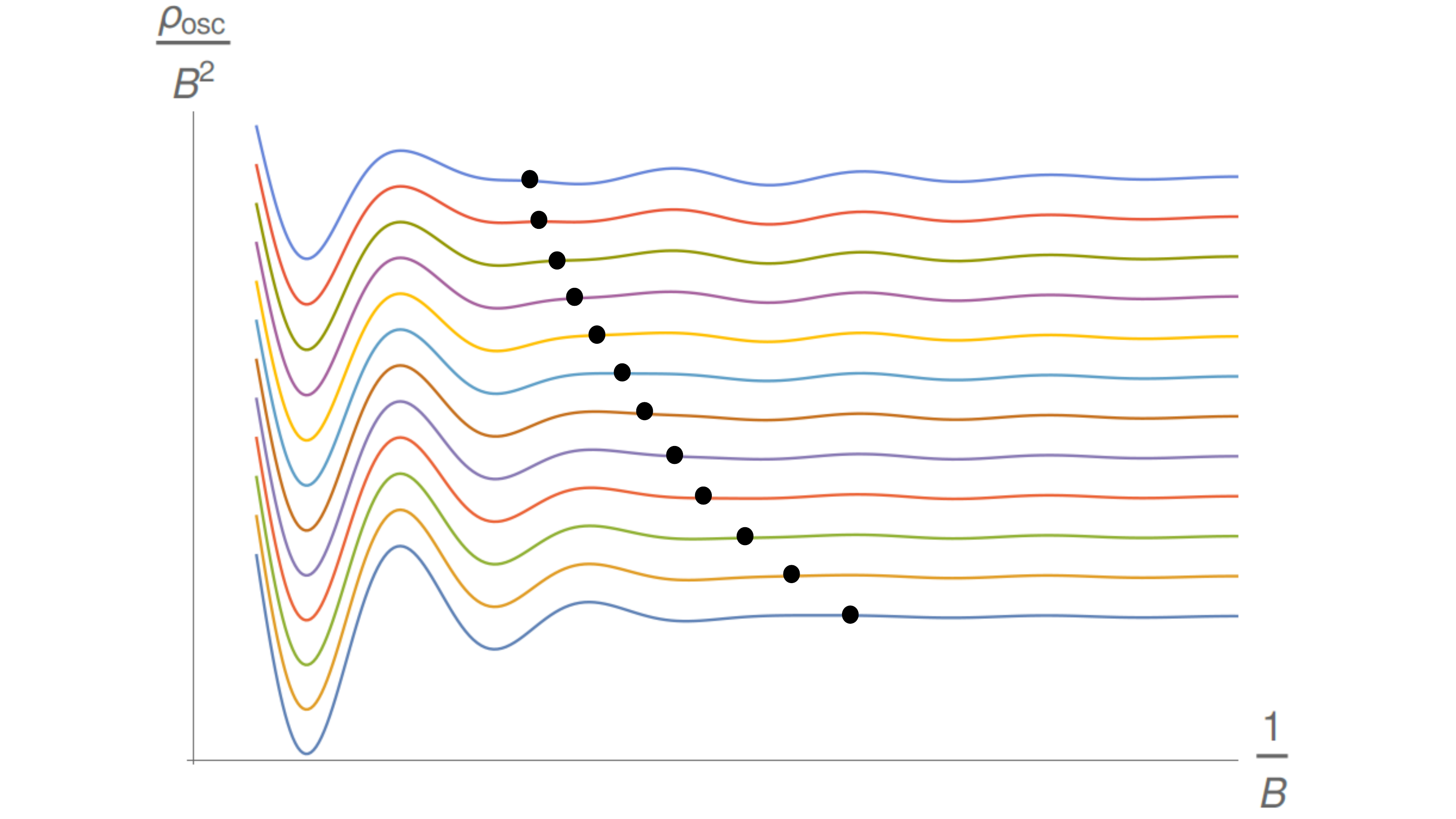}
\label{fig12_supp}}
\caption{(a) Oscillations at different temperatures obtained by numerical calculations on a lattice described in the previous section. Here $\mu=-0.09t$ where $t$ is the hopping parameter. Temperature increases in steps of $0.003t$ going up. The critical field at which the phase change happens, marked by black circles, is seen to move to the left with increase in temperature. (b) Same as in (a) but for different $\mu$ at a fixed temperature $T=0.01t$. $\mu$ varies between $-0.08t$ to $-0.091t$ in steps of $0.001t$ going up.}
\label{fig4}
\end{figure}

\section{Tight-binding calculation for numerical verification}

In order to test the predictions of our theory, we perform numerical calculations on lattice using the tight-binding method. To simulate our model, we consider two square lattices with some overlap, and hybridize them to open a gap. The Hamiltonian can be written as
\beq
H=\left[\Delta \sum_ic_i^{\dagger}c_i-\sum_{<i,j>}\left(t_{ij}c_i^{\dagger}c_j+h.c.\right)\right]+\left[-\Delta \sum_id_i^{\dagger}d_i+\sum_{<i,j>}\left(t_{ij}d_i^{\dagger}d_j+h.c.\right)\right]+\left[\zeta\sum_i\left(c_i^{\dagger}d_i+h.c\right)\right]
\label{tb}
\eeq
In the absence of a magnetic field, on putting $t_{ij}=t$, Eq.~(\ref{tb}) produces the spectrum $E^{\pm}(\mb{k})=\pm\sqrt{(\varepsilon(\mb{k})-\Delta)^2+\zeta^2}$, with $\ve(\mb{k})$ given by the usual square lattice spectrum, $\ve(\mb{k})=-2t(\mathrm{cos}k_xa+\mathrm{cos}k_yb)$, where $a$ and $b$ are lattice constants. In our calculations, we take $\Delta=2t$ to ensure that the gap opens away from half-filling for each lattice and that there is no effect from Van Hove singularities. We have chosen a symmetric structure for $H$ in Eq.~(\ref{tb}) such that a gap of $2\zeta$ opens around zero energy.

Magnetic field is introduced via Peierls substitution for the hopping parameters as $t_{ij}=te^{ie\int_i^j \mb{A}.d\mb{l}}$, where $\mb{A}$ is the magnetic vector potential, and $d\mb{l}$ denotes an infinitesimal line element from points $i$ to $j$ on the lattice. We use the gauge $\mb{A}=(0,Bx)$. Writing $x$ as $la$, where $l$ is an integer, the phase in the hopping parameter becomes $e\int A_ydy=2\pi l\phi/\phi_0$, with $\phi$ being the magnetic flux and $\phi_0$ being the flux quantum. It is seen that for $\phi/\phi_0=p/q$, where $p$ and $q$ are integers, a periodicity of $qa$ in the $x$-direction is restored. In our calculations we take $p=1$. Going to the Fourier space, we can cast the first term in square brackets in Eq.~(\ref{tb}) in terms of a $q-$component basis $C=[c^1,\cdots,c^q]$ as
\beq
-tc^{n+1}_{k_x,k_y}e^{ik_xa}-tc^{n-1}_{k_x,k_y}e^{-ik_xa}-2tc^n_{k_x,k_y}[\mathrm{cos}(k_yb-2\pi n\phi)+\Delta], \quad n=1,\cdots, q
\label{squarepart}
\eeq
Similarly for the second term in Eq.~(\ref{tb}) in terms of $D=[d^1,\cdots,d^q]$. Finally, in the basis of $C$ and $D$, one can now write the Hamiltonian as
\beq
H=\sum_{\mb{k}}
\begin{pmatrix}
C^{\dagger}_{\mb{k}}&D^{\dagger}_{\mb{k}}
\end{pmatrix}
\begin{pmatrix}
\mathcal{H}_{\mb{k}}&\mathcal{G}_{\mb{k}}\\
\mathcal{G}_{\mb{k}}&\mathcal{H}_{\mb{k}}
\end{pmatrix}
\begin{pmatrix}
C_{\mb{k}}\\
D_{\mb{k}}
\end{pmatrix}
\eeq
with $\mathcal{H}$ a $q\times q$ matrix given by (\ref{squarepart}) and $\mathcal{G}=\zeta\mathcal{I}$, where $\mathcal{I}$ is a $q\times q$ identity matrix. The problem is thus reduced to an eigenvalue problem for a $2q\times 2q$ matrix.

Solving the eigenvalue problem numerically, we get the discrete energy values for each value of $\phi=n/q$, $n=1,\cdots,q$. This allows us to calculate the grand canonical potential $\Omega$ directly, which exhibits oscillations superimposed on a smooth background. For small fields, the background is given by $\Omega_{bg}(B)\approx\Omega(0)-1/2\chi B^2$, where $\chi$ is the susceptibility. Using the method in Ref.~\cite{piechon}, we calculate $\Omega(0)$ and $\chi$ exactly from the spectrum at zero field, and subtract the resulting $\Omega_{bg}(B)$ from the numerically calculated $\Omega$, to extract quantum oscillations. The process of subtracting the background this way not only helps us to extract the oscillating part, but also provides independent proof of the accuracy of the numerical calculations.

\section{Dependence of the critical field on temperature}

When $\mu\ne 0$, Eq.~(\ref{delmunotzero}) predicts non-trivial features in the temperature dependence of oscillations. In particular, with increase in temperature, oscillations go to zero at a critical value of inverse field, $1/\omega_c^{\ast}$, suffer a phase change by $\pi$, and grow again. The critical field is a function of both $T$ and $|\mu|$. Here we show this dependence by means of numerial calculation of the density of states on the lattice using the method described in the previous section (same  behavior is obtained for the grand canonical potential). Our results are shown in Figs.~\ref{fig11_supp} and \ref{fig12_supp}. The dependence of $1/\omega_c^{\ast}$ on temperature, as shown in Fig.~\ref{fig11_supp} is useful in distinguishing whether the phase change is due to $\mu$ or topological properties since in the latter case, the critical field does not change with temperature \cite{zhasupp}. This results has been quoted in the main text. Also, to unambiguously confirm that the origin of the phase change is the asymmetry in the chemical potential, in Fig.~\ref{fig12_supp} we plot oscillations for different $\mu$ (inside the gap) at a fixed temperature. Eq.~\ref{delmunotzero} predicts that  $1/\omega_c^{\ast}$ should also shift as a function of $|\mu|$. This is confirmed in the figure.

\end{widetext}


\begin{thebibliography}{99}
\bibitem{com1}  One can argue that the presence of the artificial cutoff $\Lambda$ can give rise to spurious oscillations \cite{zha}. However, as mentioned before, the decomposition in Eq.~(\ref{omegaeh}) can as well be applied to the ungapped case when $\zeta=0$, where we are guaranteed to have no oscillations other than the ones arising from the Fermi level, i.e., the results do not depend on the cutoff. In the case of $\zeta\ne 0$, we formally choose the same cutoff, and avoid the problem of spurious oscillations.
\bibitem{supp} [URL will be inserted by publisher]
\end{thebibliography}
\begin{thebibliography}{99}
\bibitem{comment1} A more general expression would be $S^{-,e/h}(E)=S(-\zeta)+\mathrm{sgn}(e/h)[\frac{E+\zeta}{E^{(n)}(-\zeta)/n!}]^{1/n}$, where $n$ is some integer and $E^{(n)}$ is the $n$-th derivative of $E$ with respect to $S$. However, this does not lead to any qualitatively new feature.
\end{thebibliography}
\end{document}